\documentclass[prb,aps,twocolumn,floatfix]{revtex4}
\usepackage{graphicx,amsmath,amssymb}

\begin{document}

\title{Numerical studies of the Pfaffian model of the $\nu=\frac{5}{2}$ fractional quantum Hall effect}
\author{Csaba T\H oke$^{*}$}
\author{Nicolas\ Regnault$^{\dagger}$}
\author{Jainendra K.\ Jain$^{*}$}
\affiliation{$^{*}$Department of Physics, 104 Davey Laboratory, The Pennsylvania State University, Pennsylvania, 16802}
\affiliation{$^{\dagger}$ Laboratoire Pierre Aigrain, D\'epartement de Physique, 24 rue Lhomond, 75005 Paris, France}

\begin{abstract}  
The Pfaffian model has been proposed for
the fractional quantum Hall effect (FQHE) at $\nu=\frac{5}{2}$.  We examine 
it for the quasihole excitations by comparison with exact diagonalization 
results. Specifically, we consider the structure of the low-energy spectrum, 
accuracy of the microscopic wave functions, particle-hole symmetry, 
splitting of the degeneracies, and off-diagonal long range order.  We also 
review how the 5/2 FQHE can be understood without appealing to the Pfaffian 
model.  Implications for nonabelian braiding statistics will be mentioned.
\end{abstract}
\maketitle

\section{Introduction}

$5/2$ is the only even denominator fraction securely observed in a single layer 
fractional quantum Hall effect (FQHE) \cite{Willett1,Pan,Eisen02}.  
(The fraction 7/2 is trivially related to it by particle 
hole symmetry in the second Landau level.)  
The model of noninteracting composite fermions predicts a compressible Fermi sea  
at half filled {\em lowest} Landau level, which provides a good description of the compressible state here \cite{FStheory,FSexp}.
A promising scenario for the incompressible state at the half filled second Landau level 
at $\nu=5/2$ is based on the idea of pairing of composite fermions, described by a Pfaffian 
wave function\cite{Moore1,GWW1}.  Several studies have 
supported this interpretation \cite{Morf1,Park1,Scarola1,Rezayi00,Scarola2}.

The Pfaffian wave function is the exact ground state of a singular \emph{three-body} model interaction (cf.\ Eq.~\ref{modelint} below).  Exact solutions for quasiholes are also available for this model interaction.
A case has been made, both from analytical arguments\cite{Moore1,Nayak96,Read96} and numerical calculations \cite{TSimon}, that 
these Pfaffian quasiholes have the remarkable property of nonabelian braiding statistics.
Recently, the nonabelian statistics has taken additional importance because of 
proposals to test it experimentally \cite{DasSarma05,Stern06,Bonderson06a},
and to exploit it for quantum computation \cite{DasSarma05,Freedman06,Bonderson06b,Bena06}.  
That makes it important to perform an examination of the applicability of the 
Pfaffian model to the real, Coulomb solution.  The Coulomb ground state wave function 
has been compared to the Pfaffian ground state wave function in the past \cite{Morf1,Scarola2} and found to have overlaps in the range 0.69-0.87 for 8 to 16  particles.  Our recent comparisons of the Pfaffian quasiholes and the real Coulomb 
quasiholes \cite{Toke07} showed a worse agreement.  Because the 
route to non-Abelions is via the Pfaffian model and the degeneracies it 
implies, these studies have relevance to nonabelian braiding statistics as well.

This paper briefly reviews our previous work, at the same time
providing many new results 
relevant to this problem.  In Sec.~\ref{sec-review} the Pfaffian model is defined and 
some relevant results are reviewed.
In Sec.~\ref{sec-phsymm} we comment on the particle-hole symmetry violation by the Pfaffian family of states.
In Sec.~\ref{sec-odlro} the absence of off-diagonal long range order in the Pfaffian state is pointed out,
and the relevance of this finding is discussed.
In Sec.~\ref{sec-splitting} we check the assertion, commonly made in the literature, that the energy difference of the
quasihole states, that are degenerate for the three-body model interaction, remains exponentially small for the Coulomb interaction.
In Sec.~\ref{sec-separate} we attempt to separate the postulated charge $\frac{1}{4}$ quasiholes for
the model interaction as well as the Coulomb interaction.
Finally, in Sec.~\ref{sec-alternative} we elaborate an alternative approach for 
the explanation of the $\frac{5}{2}$ FQHE.  Short reports on parts of this 
paper have appeared elsewhere \cite{Toke06,Toke07}.


\section{The Pfaffian model}
\label{sec-review}

Throughout this article we will assume that the lowest Landau level (LL) is full and inert, and the
two-dimensional electron gas in the second LL is fully polarized. All calculations 
are performed in the spherical geometry.  The objective is to determine
the ground state and the low-energy excitations for the Coulomb interaction
\begin{equation}
V^{\rm (C)}=\frac{e^2}{\epsilon}\sum_{i<j}\frac{1}{\left|\mathbf r_{i}-\mathbf r_{j}\right|},\index{$H_I$}
\end{equation}
in second LL at filling factor $\nu=1/2$ ($\epsilon$ is the static dielectric constant of the host semiconductor).  This problem is equivalent to electrons in the lowest 
Landau level interacting with an effective interaction $V^{\rm eff}$.  We will use 
the lowest LL to simulate the second LL physics in what follows.

The ``Pfaffian model" considers a three-body model interaction\cite{GWW1,Read96}, which in the spherical geometry takes the form
\begin{equation}
V^{\rm Pf}=\frac{e^2}{\epsilon l_B} \sum_{i<j<k} P_{ijk}(L_{\rm max})
\label{modelint}
\end{equation}
where $P_{ijk}(L_{\rm max})$ is the projection operator onto an
electron triplet with orbital angular momentum $L_{\rm max}=3Q-3$.
The angular momentum $L_{\rm max}$ corresponds to the closest possible configuration of an electron triplet.  Thus, 
$V^{\rm Pf}$ does not penalize the closest approach of \emph{two} electrons, but
there is an energy cost when \emph{three} electrons are in their closest configuration.

This model has a unique, zero energy ground state at $\nu=1/2$
 (Moore and Read\cite{Moore1}):
\begin{equation}
\label{pfaffdisk}
\Psi^{\rm Pf}_0=\text{Pf}\left(\frac{1}{z_i-z_j}\right)\Phi_1^2,\quad\quad\Phi_1=\prod_{i<j}(z_i-z_j),
\end{equation}
where ``Pf" refers to ``Pfaffian." This wave function describes a paired state of composite fermions. This model also produces exact zero energy eigenfunctions for 
quasiholes, as the flux through the sphere is increased.  These zero energy 
eigenstates are referred to as the 
``Pfaffian quasihole (PfQH) states."  As $[V^{\rm Pf},L^2]=0$, the states spanning the PfQH sector may be chosen with a definite orbital angular momentum $L$.
(See Ref.~\onlinecite{Read96} for a thorough study of the PfQH sector on the sphere.)
Appropriate linear combinations of these states produce spatially localized quasiholes. 
For two quasiholes at $\eta_1$ and $\eta_2$, the wave function is given by\cite{Moore1}
\begin{equation}
\Psi^{\rm Pf}_{\rm 2-qh}={\rm Pf}\left(\frac{(z_i-\eta_1)(z_j-\eta_2)+(i\leftrightarrow j)}{(z_i-z_j)}\right)\Phi_1^2.
\label{pfaff2holes}
\end{equation} 
For two coincident quasiholes, $\eta_1=\eta_2\equiv\eta$, this reduces to a charge $\frac{1}{2}$ vortex:
\begin{equation}
\label{pfaffvortex}
\Psi_V=\prod_i(z_i-\eta)\Psi^{\rm Pf}_0. 
\end{equation}
Separately, each quasihole has a charge deficiency of $\frac{1}{4}$ associated with it.
Unlike for the vortex, the density does not vanish at the center of a quasihole.
Analogous wave functions can be written for an even number ($2m$) of quasiholes.
Exact wave functions for quasiparticles are not available.

Several wave functions can be associated for a given configuration of $2m$ 
quasiholes, which correspond,
in the appropriate generalization of Eq.~(\ref{pfaff2holes}) to $2m$ quasiholes,  to different ways of grouping  
half of the $\eta_k$'s with $z_i$ and the other half with $z_j$.
It has been shown\cite{Nayak96} that only $2^{m-1}$ of these functions are linearly independent.
Adiabatic braiding of quasiholes (which is feasible for a gapped system) can take the system from
one linear combinations of PfQH states to another, which lies at the origin of nonabelian statistics of quasiholes.

To study bulk properties, it is convenient to formulate the problem of
interacting electrons in the spherical geometry, in which the electrons move on the
surface of a sphere and a radial magnetic field is produced by
a magnetic monopole of strength $Q$ at the center.\cite{Haldane83,Fano}
Here $2Q\phi_0$ is the magnetic flux through the surface of the sphere; $\phi_0=hc/e$, and
$2Q$ is an integer by Dirac's quantization condition.
Then wave functions in Eqs.~(\ref{pfaffdisk}-\ref{pfaffvortex}), which are written 
for the disk geometry, 
can be mapped to the sphere by the stereographic mapping\cite{Fano}, which amounts to the substitution
\begin{equation}
(z_a-z_b)\to(u_av_b-v_au_b),
\end{equation}
for all coordinate differences, where $u_a=\cos\frac{\theta_a}{2}e^{-i\phi_a/2}$ and
$v_a=\sin\frac{\theta_a}{2}e^{i\phi_a/2}$ are spinorial coordinates on the sphere.
The orbital angular momentum quantum number is denoted by $L$.


\section{Particle-hole symmetry}
\label{sec-phsymm}

The exact Coulomb eigenstates in any given Landau level satisfy particle-hole 
symmetry, i.e., the exact eigenstates at $\nu$ and $1-\nu$ are related by 
particle-hole transformation.  The wave functions in the CF theory\cite{Jain1} 
satisfy particle 
hole symmetry to a very good approximation, 
even though there is no symmetry principle that so requires.  For example, the wave functions 
at $\nu=n/(2n-1)$, given by $\Psi_{n/(2n-1)}={\cal P}_{\rm LLL}\Phi_1^2[\Phi_n]^*$, 
are almost identical to the those obtained by particle-hole transformation of the wave functions $\Psi_{n'/(2n'+1)}={\cal P}_{\rm LLL}\Phi_1^2\Phi_{n'}$, with $n=n'+1$.  

\begin{figure}[!htbp]
\begin{center}
\includegraphics[width=\columnwidth]{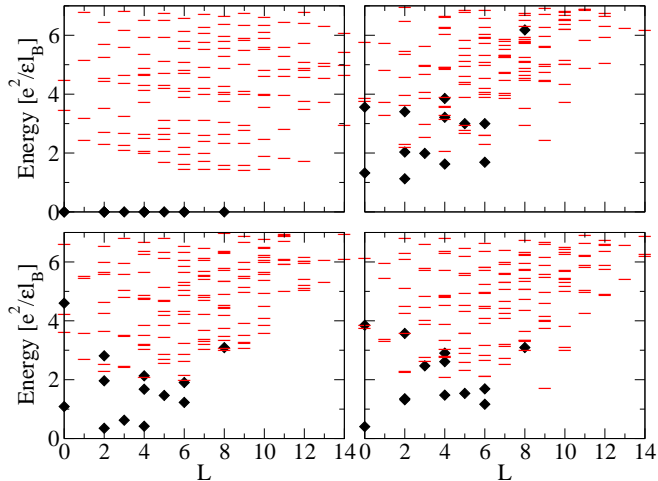}
\end{center}
\caption{\label{phsymm}
Upper left panel shows the original spectrum of $V^{\rm Pf}$ with $N=8,2Q=2N-1$ (four quasiholes); the Pfaffian quasihole states have zero energy.  Also shown are the 
spectra for the p-h conjugate states (top right),
the p-h symmetrized states (bottom left), and p-h antisymmetrized states (bottom right).
The diamonds show the states derived from the Pfaffian quasihole branch.
}
\end{figure}

The three body interaction does not satisfy particle-hole (p-h) symmetry.  
To get a feel for 
the extent to which this symmetry is broken, we have considered the system of 
$N=8$ particles at $2Q=15$.  In this case, particle hole transformation gives eight 
holes (to be distinguished from {\em quasi}holes)
at $2Q=15$.  We obtain the exact spectrum of 
the $V^{\rm Pf}$ model interaction, which is given in the upper left panel of 
Fig. \ref{phsymm}.  This system corresponds to four quasiholes, and has a number of 
zero energy states, which form the Pfaffian quasihole sector.  
We obtain the particle-hole conjugate of each eigenstate, called $\Psi^c$, and calculate  
its energy expectation value for the $V^{\rm Pf}$ interaction. When there are several degenerate Pfaffian quasihole states, we diagonalize $V^{\rm Pf}$ in the 
subspace of the p-h conjugate states to obtain the energies.
The resulting spectrum 
is shown in the top right column of Fig. \ref{phsymm}. (For the Coulomb interaction, 
this exercise would produce a spectrum identical to the original one, 
apart from an overall energy shift.)  We construct 
symmetrized states $\Psi^s\propto(\Psi+\Psi^c)$, which satisfy particle-hole 
symmetry by construction; the resulting spectrum 
for these states is given in the lower left panel of Fig. \ref{phsymm}.  The lower 
right spectrum is for antisymmetrized states $\Psi^a\propto(\Psi-\Psi^c)$.  

Table \ref{phsymm} shows the squared overlaps between the original Pfaffian quasihole 
states with the various states obtained with the help of p-h conjugation.
To handle the multiplicity of the PfQH sector for $L=0,2,4,6$ (cf.\ Fig.~\ref{phsymm}),
the overlap between two subspaces has been defined in a basis-independent 
manner (see caption of Table \ref{phsymm}).
The overlaps are not particularly high; for example, in the $L=0$ part of the quasihole branch, which contains two states for $N=8$, the overlap is 0.511, and 
deteriorates for higher $L$'s. Similar numbers are obtained for other states in the PfQH sector.
The near orthogonality of $\Psi$ and $\Psi^c$ at $L=8$ is accompanied by a very high energy of $\Psi^c$.

These results demonstrate a substantial breakdown of
the p-h symmetry by the $V^{\rm Pf}$ 
interaction.  The Pfaffian quasihole band is absent in all of the new spectra;  the states 
derived from the Pfaffian quasihole band are mixed up with other states. 
 It has been shown\cite{Rezayi00} that the particle-hole symmetrization ($\Psi\to\Psi^s$) of the Pfaffian wave function improves the
overlap with the Coulomb ground state. Our results show, however, that this also destroys the degeneracy of PfQH sector.  One 
can ask whether the nonabelian statistics of the Pfaffian quasiholes is 
robust to p-h symmetrization; we are not able resolve this question definitively 
by a direct calculation of the braiding phases, which requires much larger systems.

\begin{table}[htb]
\begin{center}
\begin{center}
\begin{tabular}{l|l|l|l}
\hline\hline
         & p-h & symmetric & antisymmetric  \\
         & conjugate & combination & combination\\
\hline
$O(L=0)$ & 0.511 & 0.425 & 0.575 \\
$O(L=2)$ & 0.431 & 0.542 & 0.458 \\
$O(L=3)$ & 0.357 & 0.798 & 0.201 \\
$O(L=4)$ & 0.255 & 0.641 & 0.359 \\
$O(L=5)$ & 0.001 & 0.511 & 0.489 \\
$O(L=6)$ & 0.233 & 0.443 & 0.557 \\
$O(L=8)$ & $4\times10^{-7}$ & 0.500 & 0.500 \\
\hline\hline
\end{tabular}
\end{center}
\end{center}
\caption{\label{phoverlap}
Squared overlaps between the subspaces spanned by the zero-energy states and the subspaces spanned by their particle-hole conjugate,
particle-hole symmetrized, and particle-hole antisymmetrized images, respectively.
Squared overlaps are defined as
${\cal O}=\sum_{i,j}^{\cal N}|\langle\Psi_{\rm 4-qh, i}|\Psi'_{\rm 4-qh, j}\rangle|^2 / {\cal N}$, where 
${\cal N}$ is the number of degenerate multiplets\cite{Read96} of $V^{\rm Pf}$ at $L$, and $i,j=1,\cdots, {\cal N}$.  
}
\end{table}


\section{Off-diagonal long range order}
\label{sec-odlro}

We wish to stress that the Pfaffian wave function does not represent a 
true superconductor;  the pairing of composite fermions 
opens a gap to produce FQHE but does not establish long range 
phase coherence in the electronic state.  For this purpose, we 
calculate the off-diagonal long range order parameter:
\begin{equation}
|G(\mathbf r_1,\mathbf r_2,\mathbf r_1',\mathbf r_2')|=
\langle \Psi_0|\hat\psi^\dagger(\mathbf r_1')\hat\psi^\dagger(\mathbf r_2')
\hat\psi(\mathbf r_2)\hat\psi(\mathbf r_2)|\Psi_0\rangle,
\end{equation}
where $\hat\psi(\mathbf r)$ and $\hat\psi^\dagger(\mathbf r)$ are the usual 
annihilation and creation field operators. We place the primed coordinates near 
the north pole, separated by a distance equal to the magnetic length, and the 
unprimed coordinates at the south pole, also separated by a distance equal to the magnetic 
length. The results in Table \ref{odlro}, obtained by Monte Carlo 
calculation, demonstrate the absence of off-diagonal long-range order
in the Pfaffian wave function.

\begin{table}[htb]
\begin{center}
\begin{tabular}{r|c}
\hline\hline
$N$ & $G(\mathbf r_1,\mathbf r_2\mathbf ,r_1',\mathbf r_2')$ \\
\hline
4  & 0.0005(9)\\
6  & 0.001(2) \\
8  & 0.0000(1)\\
10 & 0.0002(5)\\
\hline\hline
\end{tabular}
\end{center}
\caption{\label{odlro}
Off-diagonal long-range order parameter $G(\mathbf r_1,\mathbf r_2\mathbf ,r_1',\mathbf r_2')$ with $\mathbf r_1$ and $\mathbf r_2$ separated by $l_B$ about the north pole, and
$\mathbf r_1'$ and $\mathbf r_2'$ separated by $ l_B$ about the south pole for the paired CF wave function $\textrm{Pf}(1/(z_i-z_j))\Phi_1^2$.
}
\end{table}

\section{Testing the Pfaffian quasihole wave function}

\begin{figure}[htbp]
\begin{center}
\includegraphics[width=\columnwidth]{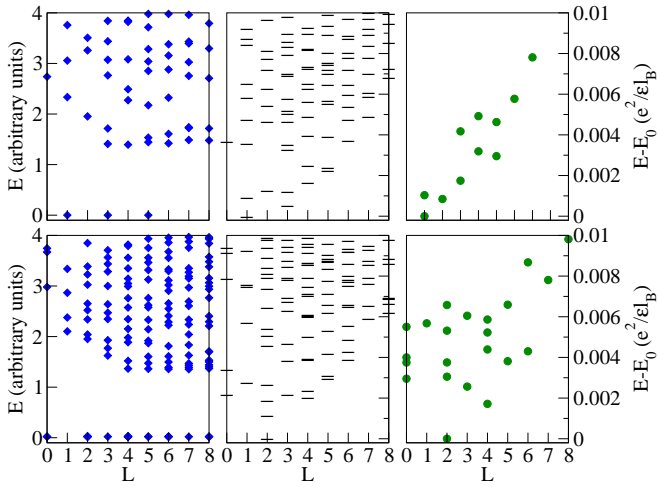}
\end{center}
\caption{\label{comparison}
Spectra at $\nu=\frac{5}{2}$ for the model interaction $V^{\rm Pf}$ (left column), 
and the Coulomb interaction (right column) for $N=10$ particles at $2l=18$ 
(top row) and $2l=19$ (bottom row).  For the $V^{\rm Pf}$ interaction,
two (four) quasiholes are expected for $2l=18$ ($2l=19$).
The spectra on the left were also given in Ref.~\onlinecite{Read96}. This figure 
is taken from Ref. \onlinecite{Toke06}.}
\end{figure}

\begin{figure}[!htbp]
\begin{center}
\includegraphics[width=\columnwidth]{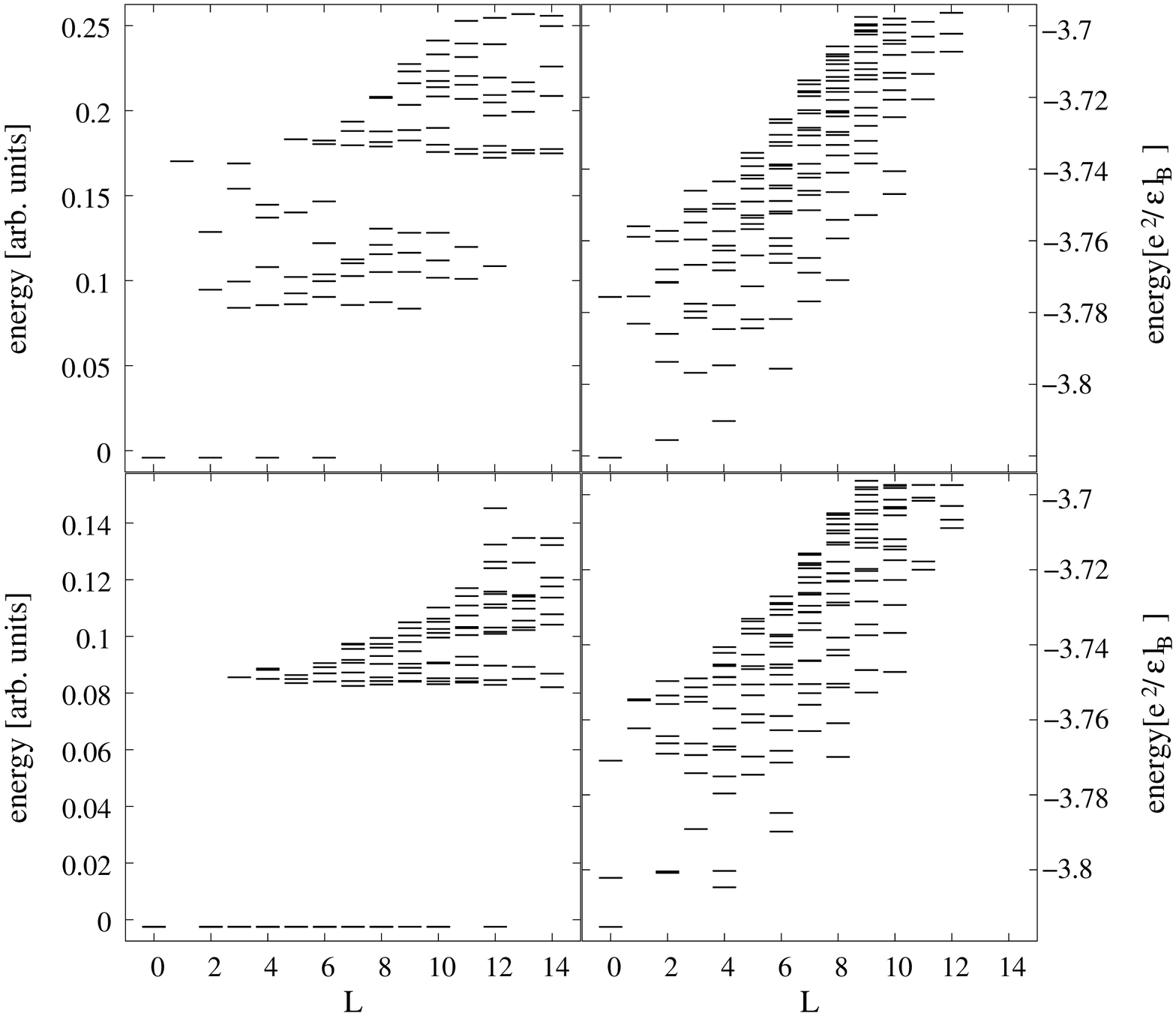}
\end{center}
\caption{\label{n12}
Spectra at $\nu=\frac{5}{2}$ for the model interaction $V^{\rm Pf}$ (left column), 
and the second Landau level Coulomb interaction (right column) for $N=12$ particles for 
two (upper row) and four (lower row) quasiholes.}
\end{figure}

\begin{figure}[!htbp]
\begin{center}
\includegraphics[width=\columnwidth]{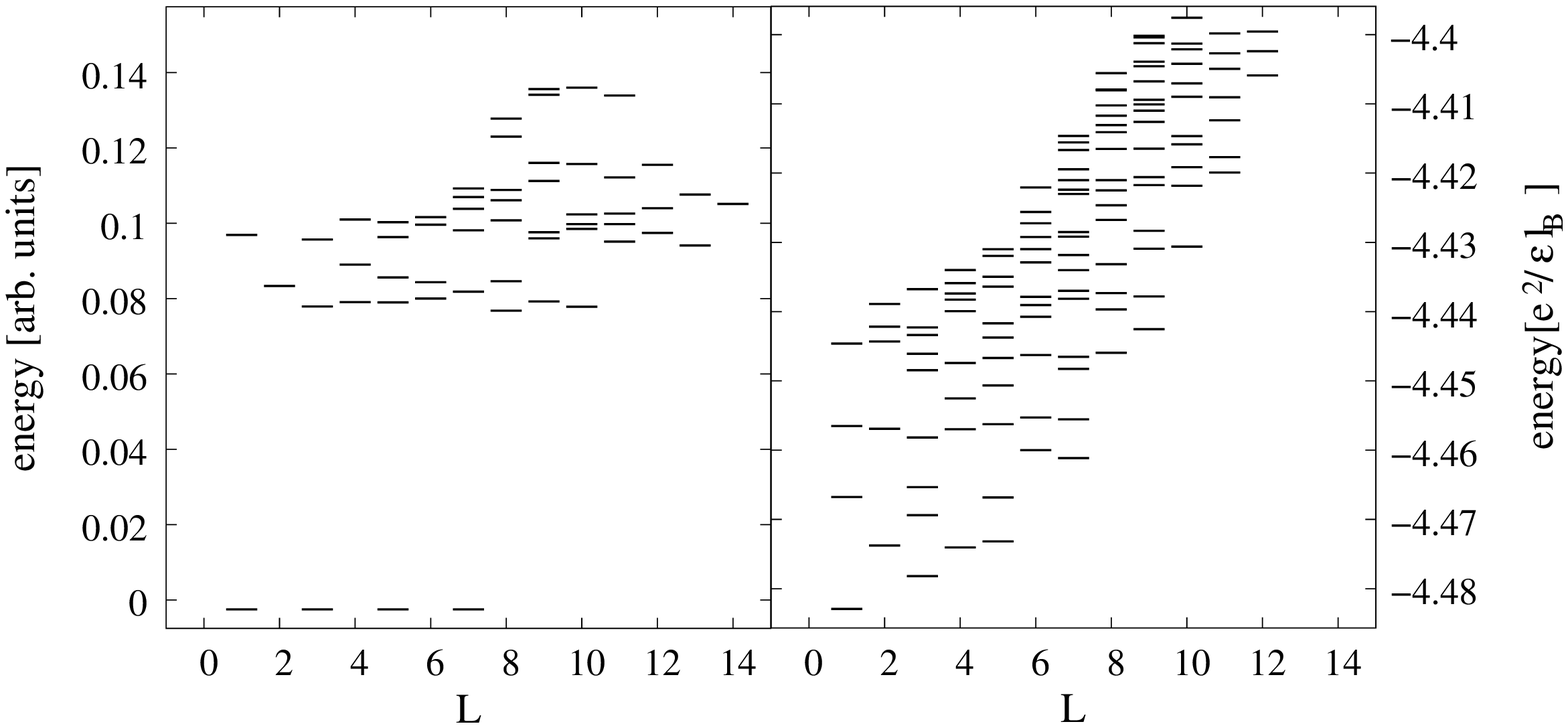}
\end{center}
\caption{\label{n14}
Spectra at $\nu=\frac{5}{2}$ for the model interaction $V^{\rm Pf}$ (left column), 
and the second Landau level Coulomb interaction (right column) for $N=14$ particles for 
two quasiholes. 
}
\end{figure}

We have recently carried out comparisons between the Pfaffian and Coulomb 
quasiholes \cite{Toke07}.  Figs. \ref{comparison}, \ref{n12}, and \ref{n14} show the 
spectra for states with two and four quasiholes for $N=10$ and 12 electrons.  For 
10 electrons, the Pfaffian model predicts zero energy states at $L=1,3,5$ and
$L=0^2, 1^0, 2^4, 3^1, 4^4, 5^2, 6^3, 7^1, 8^2, 9^0, 10^1$, respectively
(the superscript denotes the degeneracy), for two and four quasiholes.  
These states form the Pfaffian quasihole band.  For 12 electrons, 
the Pfaffian quasihole band contains states at $L=0,2,4,6$ for two quasiholes 
and $L=0^3, 1^0, 2^4, 3^2, 4^5, 5^2, 6^5, 7^2, 8^3, 9^1, 10^2,11^0,12^1$ for four quasiholes.  For 14 electrons, 
the Pfaffian quasihole band for two quasiholes has states at angular momenta $L=1, 3, 5,7$.

\begin{figure}[!htbp]
\begin{center}
\includegraphics[width=\columnwidth, keepaspectratio]{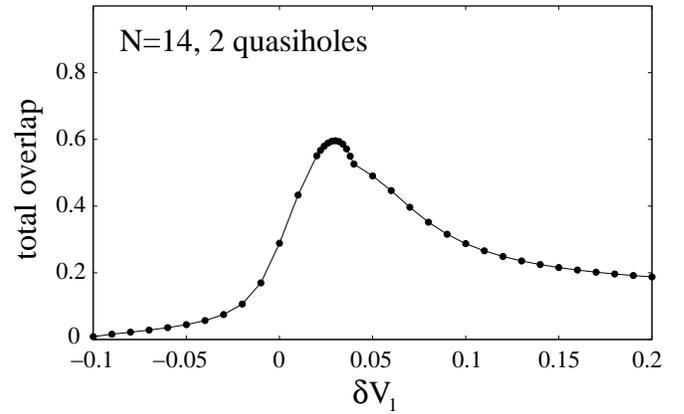}
\end{center}
\caption{\label{qhoverlap14}
The overlap between the low-energy excitations of the second-LL 
Coulomb and $V^{\rm Pf}$ interactions for $N=14$ particles as the
leading pseudopotential $V_1$ is changed.}
\end{figure}

The Coulomb spectra in Figs. \ref{comparison}-\ref{n14} do not show well 
defined bands that have a one-to-one correspondence with the Pfaffian 
quasihole bands.
Ref. \onlinecite{Toke07} gives overlaps between the Pfaffian and Coulomb quasihole 
states, which are generally worse than for the ground state.  Fig. \ref{qhoverlap14}
depicts for the two quasihole state (for 14 electrons) 
the ``total overlap," defined as ${\cal O}=\sum_{L=1,3,5,7}|\langle\Psi_{\rm 2-qh}^{L}|\Psi_{\rm coul}^{L}\rangle|^2 / 4$ where $|\Psi_{\rm 2-qh}^{L}\rangle$ is the two quasihole state with $L_z=L$ and  $|\Psi_{\rm coul}^{L}\rangle$ is the Coulomb ground state with $L_z=L$.  This figure shows the dependence of the overlap on the 
form of the interaction; by increasing the $V_1$ pseudopotential of the Coulomb interaction by 0.03 units, it is possible to increase the overlap from 0.3 to 0.6.
For large $\delta V_1$, the solution is essentially the lowest-LL 
Coulomb solution; Fig. \ref{qhoverlap14} thus shows that the Pfaffian 
wave functions provide a comparable description of the state in the lowest 
two Landau levels.

\section{Energy splitting of the Pfaffian quasihole states}
\label{sec-splitting}

The Pfaffian model predicts a $2^{m-1}$ degenerate wave functions for any given 
configuration of $2m$ quasiholes, which is 
responsible for the emergence of nonabelian braiding statistics.
Any deviation from the model interaction $V^{\rm Pf}$ lifts this degeneracy.
but a case can be made that if the energy splitting of these states remains exponentially
small as a function of the distance between the quasiholes, the idea of nonabelian statistics remains experimentally relevant.
It would be of interest to test how the splitting behaves in a realistic calculation.
Unfortunately, a good model for the Coulomb quasiholes is not available, 
and it is not known how separated quasiholes can be produced in exact diagonalization 
studies \cite{Toke07} (Section VII).  We study how the Coulomb interaction splits the 
degeneracy while restricting to the PfQH sector.  In light of the above comparisons,  such a restriction is not necessarily a valid approximation, 
because the Coulomb interaction 
causes a substantial mixing with states outside of the PfQH sector.  However, 
a more accurate calculation is currently not feasible.

The calculation requires at least four quasiholes, which we place on the sphere at maximal separation, i.e. at the vertices of a regular tetrahedron.
The Coulomb interaction in the first and second LLs is diagonalized in the space spanned by two Pfaffian quasihole wave functions.  The overlap and interaction matrices 
are calculated by Monte Carlo methods; an orthonormal basis is 
found by the standard Gram-Schmidt procedure; and the interaction is  
diagonalized in this basis.  The Coulomb interaction in the second LL is  
simulated in the lowest LL by an effective interaction of the form
\begin{equation}
\label{form}
V^{\text{eff}}(r)=\frac{1}{r}+\sum_{i=0}^M c_i r^i,
\end{equation}
where the coefficients $c_i$ are fixed so that the lowest LL pseudopotentials\cite{Haldane83} of $V^{\text{eff}}(r)$
reproduce \emph{all} of the second LL Coulomb pseudopotentials $V^{(1)}_m$ for odd integral values of $m$.
(For relevant formulas, see Ref.~\onlinecite{Toke06}.)

\begin{figure}[htb]
\begin{center}
\includegraphics[width=\columnwidth]{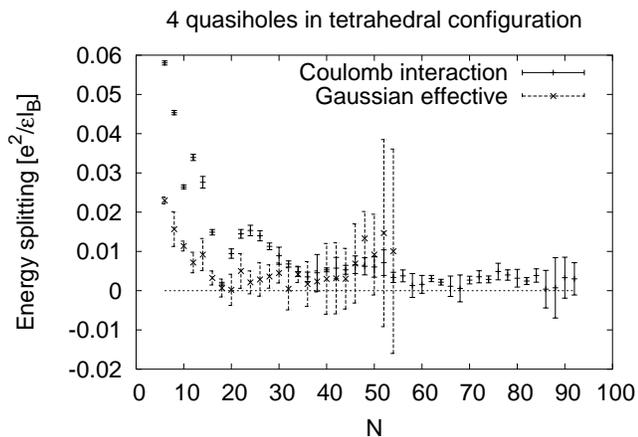}
\end{center}
\caption{\label{splitting}The energy splitting of the two four-quasihole 
wave functions for Coulomb interaction.
}
\end{figure}

\begin{figure}[htb]
\begin{center}
\includegraphics[width=\columnwidth]{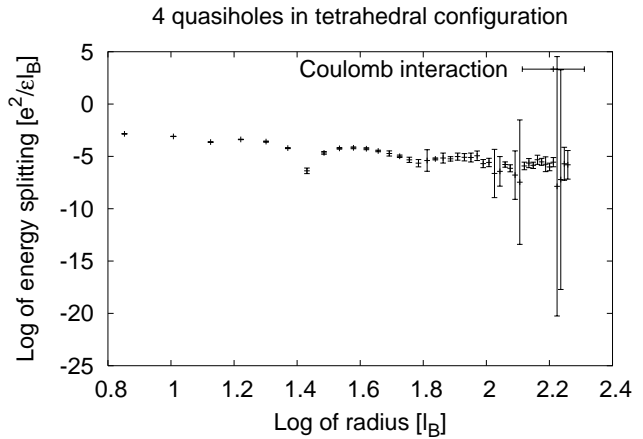}
\end{center}
\caption{\label{loglog}Same as Fig.~\ref{splitting}, with both scales logarithmic.
See line fitting on Figure \ref{fitting}.}
\end{figure}

As apparent in Fig.~\ref{splitting}, the lowest LL Coulomb interaction and the effective second LL
interaction give different results for small ($N\le 30$) systems.
Because the energy splittings are very close in the $30<N\le 54$ range,
we study larger systems ($N>54$) with the lowest LL Coulomb interaction only.
It is likely that the long distance behavior of the splitting does not depend on 
the Landau level index (given that the interaction 
at long distances is independent of the LL index).  Fig.~\ref{loglog} shows 
the lowest LL splitting.

\begin{figure}[htb]
\begin{center}
\includegraphics[width=\columnwidth]{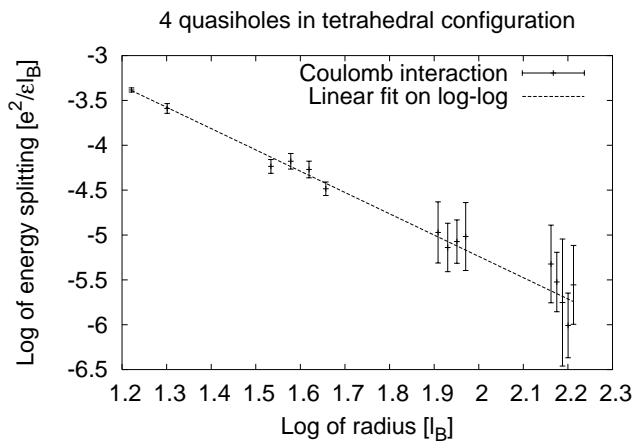}
\end{center}
\caption{\label{fitting}Line fitting on the log-log graph of the energy splitting as a function of the distance.
A straight line fitted on the local maxima of the data is consistent with a power law decay with exponent $\alpha=-2.37(6)$.}
\end{figure}

The energy splitting is a nonmonotonic function of $N$ (or $R$).
Near the local minima the error in the logarithm of the energy splitting 
is seen to become very large.  We therefore ask how the value  
of the ODLRO parameter at the local maxima decays with distance.
While inconclusive, our results are most consistent with a power law decay of the splitting: 
A straight line fits at all the four bumps in the log-log plot (Fig.~\ref{fitting}),
but not in the semilog plot (not shown).  A study of larger numbers of 
particles will be required for further confirmation, which is impractical at this 
stage, but assuming a power law, the energy splitting decays with an 
exponent $\alpha=-2.37(6)$. We stress again that the fact that the Coulomb 
interaction causes a substantial mixing with the non-PfQH sector  
diminishes the value of the calculation presented in this section.

\section{Separating quasiholes}
\label{sec-separate}

For the purpose of braiding statistics it is necessary to 
consider spatially localized states of quasiholes.  In Ref. \onlinecite{Toke07} we 
have studied states of two quasiholes in the presence of delta function 
impurities that attract the quasiholes.  We take the impurities to be 
placed at one or both of the poles so the eigenfunctions have a well defined $L_z$ 
(although they do not have a well defined $L$ quantum number).  We also 
assume sufficiently weak strengths for the impurity potential, so they do 
not cause a mixing of the Pfaffian quasihole states with higher energy 
states.  Our principal results are as follows.

\begin{figure}[!htbp]
\begin{center}
\includegraphics[width=\columnwidth, keepaspectratio]{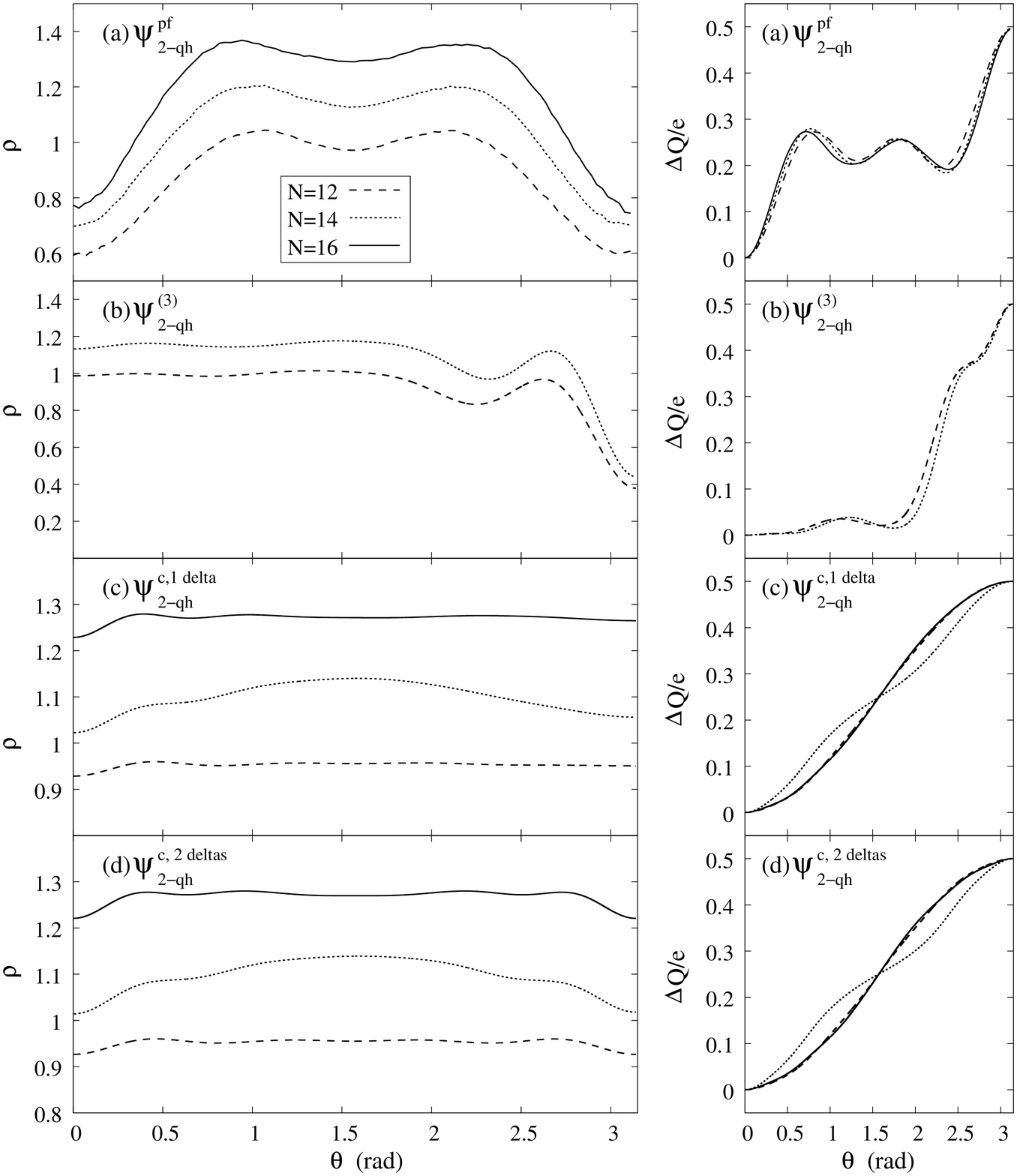}
\end{center}
\caption{Left panel: Charge densities of two quasihole states 
for (a) the Pfaffian 
wave function with two quasiholes at two poles; (b) the ground state of 
$V^{\rm Pf}$ with two delta function impurities; (c) the Coulomb ground state with one 
delta function impurity; (d) the Coulomb ground state with two impurities.
The impurities are placed on the two poles (or one pole in case of a single 
impurity), so the eigenstates have a well defined $L_z$. 
The results are for $N=12$ (dashed lines), $N=14$ (dotted lines)
and $N=16$ (solid lines) electrons at $2Q=2N-2$.  The density in (a) is 
calculated by Monte Carlo, and in other panels from exact diagonalizaion.
When the ground state has $L_z\neq 0$, there are 
two degenerate states at $\pm L_z$; we have shown only one of them for simplicity.
The normalization is chosen to ensure that the integrated density equals $N$.
Right panel: the integrated excess charge for each density, normalized so that the 
total charge excess is $\frac{1}{2}$.  Two spatially separated 
quasiholes will exhibit a step at charge $\frac{1}{4}$, as 
approximately seen in the top panel.
This figure is taken from Ref. \onlinecite{Toke07}.}\label{density2qh}
\end{figure}

For the $V^{\rm Pf}$ model a single delta function impurity 
in the lowest LL localizes a vortex (which is a combination of two quasiholes) 
rather than a single Pfaffian quasihole for the following reason.  The energy 
of a given wave function is equal to a properly weighted average of the densities at 
the positions of the delta functions (for weak impurity strengths).  
For a delta function at $(U,V)$, the lowest 
energy state (which has zero energy independent of the strength of the delta 
impurity) is the one in which {\em both} quasiholes  
localize at $(U,V)$, producing a vortex $\Psi_V$ with vanishing density at $(U,V)$.
Surprisingly,  as seen in Fig.~\ref{density2qh}(b), even two delta impurities 
fail to separate two quasiholes, even though the systems are 
sufficiently large at least for the charge-1/4 {\em Pfaffian} quasiholes to be 
well separated (top panels)

For two Coulomb quasiholes,  at first sight, one may expect that even a single 
delta function should produce well separated quasiholes, 
because it can bind one of them, which then should repel the other.  As seen 
in Fig.\ \ref{density2qh}(c,d), neither one nor 
two delta functions produce separated 
quasiholes.  In fact, the charge profile is practically identical for 
the two cases.  The situation is more 
restrictive for the Coulomb interaction because, instead of many degenerate states, 
we have a single ground state multiplet with a definite $L$.  All that {\em weak} 
disorder can do is cause a mixing between the different $L_z$ components 
of the ground state multiplet.  For the case of 
two delta functions at the two poles, $L_z$ is a good quantum number, so the 
delta functions only lift the degeneracy of the $L_z$ states.  
The lack of quasihole separation in space 
is attributable to the fact that the ground state now has a more or less 
definite $L$.  The absence of exact degeneracy
inhibits quasihole localization.

\section{Implications for braiding statistics}

The Pfaffian quasiholes are believed to obey nonabelian braiding 
statistics.  Our finite system studies of the Coulomb solutions 
do not provide a clear confirmation of the 
Pfaffian model,  and therefore of the nonabelian statistics.
We cannot rule out the possibility that 
the Pfaffian physics will be recovered in the thermodynamic limit. 
It is useful to recall, in this context, how the fractional abelian braiding statistics \cite{LM,Halperin84,Arovas} of 
the quasiparticles of the $\nu=n/(2n+1)$ states has been confirmed 
theoretically.  There, the CF theory provides a qualitatively valid description the 
quasiparticle band, as well as accurate wave functions.  These wave functions 
are then used for large systems to establish the abelian statistics \cite{Kjonsberg,Jeon03}.  These calculations also demonstrate that the braiding 
statistics is not well defined when quasiparticles are overlapping, which is why 
its confirmation requires large systems.  The nonavailability of accurate wave functions for the 5/2 quasiparticles or quasiholes prevents similar calculations 
of their braiding properties.

\section{An alternative approach for 5/2 FQHE}
\label{sec-alternative}

It is not known how the Pfaffian 
wave functions can be improved for the two body Coulomb interaction, 
due to lack of variational parameters. 
Further, the pairing of composite fermions is viewed as arising from an 
instability of the CF Fermi sea\cite{GWW1,GWW2,Scarola1}, but the CF Fermi sea 
is not a limiting case of the Pfaffian wave function.  
These observations have motivated us to approach the 5/2 FQHE from the 
CF Fermi sea end, without assuming any pairing at the outset \cite{Toke06}.  The idea is 
straightforward.  We know that noninteracting composite fermions do not show 
FQHE at 5/2; our approach is to include the residual interactions between them by constructing a basis of  ``noninteracting," or the ``unperturbed," 
CF ground and excited states and rediagonalizing the 
Coulomb interaction in that subspace to obtain the spectrum for ``interacting" 
composite fermions.  This is known as the CF diagonalization, and 
the relevant techniques are described in the literature  
\cite{JainKamilla,Mandal00}.  As usual, we simulate the second LL physics 
in the lowest LL by working with an appropriate effective interaction. 
We work at the same flux value as the Pfaffian wave function, but 
because of a technical reason \cite{Toke06}, we 
work with holes, rather than electrons.  (Holes are not to be confused with 
quasiholes.)  By particle hole symmetry, the number of holes is given by 
$N_h=(2Q+1)-N=N-2$ at $2Q=2N-3$.  In what follows, composite 
fermions are made by attaching vortices to holes rather than electrons.
We show results at ``zeroth order" CF diagonalization 
(when only the lowest energy unperturbed 
states are considered) and ``first order" CF diagonalization 
(which also includes states with one higher unit of ``kinetic energy").  
The composite fermion kinetic energy levels are called $\Lambda$ levels.

\begin{figure*}[htbp]
\begin{center}
\includegraphics[width=\textwidth]{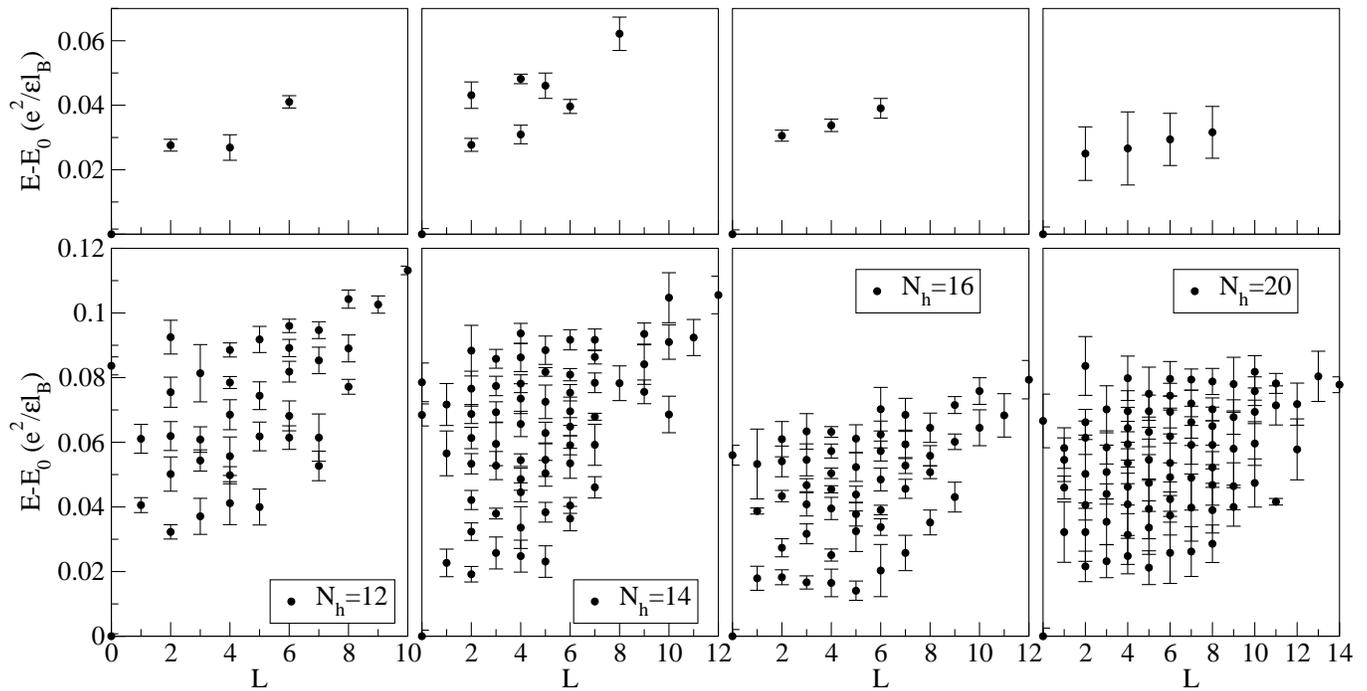}
\end{center}
\caption{\label{specrest}
Zeroth-order (top) and first-order (bottom) CF diagonalization excitation spectra for $N_h=12,14,16,20$ holes
in the second Landau level.
}
\end{figure*}

Figure \ref{specrest} shows the excitation spectra at the half filled {\em second} 
LL for $N_h=12$, 14, 16 and 20 
obtained by CF diagonalization at the zeroth and the first orders.  
($N_h=18$ is not considered as it aliases with $\nu=\frac{3}{7}$ of holes.)
The residual interaction between 
composite fermions lifts the degeneracy between various states to produce 
an incompressible state already at the lowest (zeroth) order, which neglects 
$\Lambda$-level mixing.  Although the energy gaps change by up to 
50\% in going from the the zeroth to the first order, the incompressibility is  
preserved, indicating that 
while $\Lambda$-level mixing renormalizes composite fermions, it 
does not cause any phase transition.  The overestimation of gaps at the zeroth 
order may be attributed to the very small dimensions of the CF basis. All
CF basis states are perturbations of the noninteracting CF Fermi sea, 
making it explicit that a rearrangement of composite fermions near the 
CF Fermi level is responsible for the $\frac{5}{2}$ FQHE.
Although there is some ambiguity as to which excitation  
is to be identified with the transport gap (corresponding to a far separated 
quasiparticle-quasihole pair),
an inspection indicates a gap of $\sim 0.02$, which is 
consistent with the earlier results from exact 
diagonalization\cite{Morf1,Morf2}.

\begin{figure*}[htbp]
\begin{center}
\includegraphics[width=\textwidth]{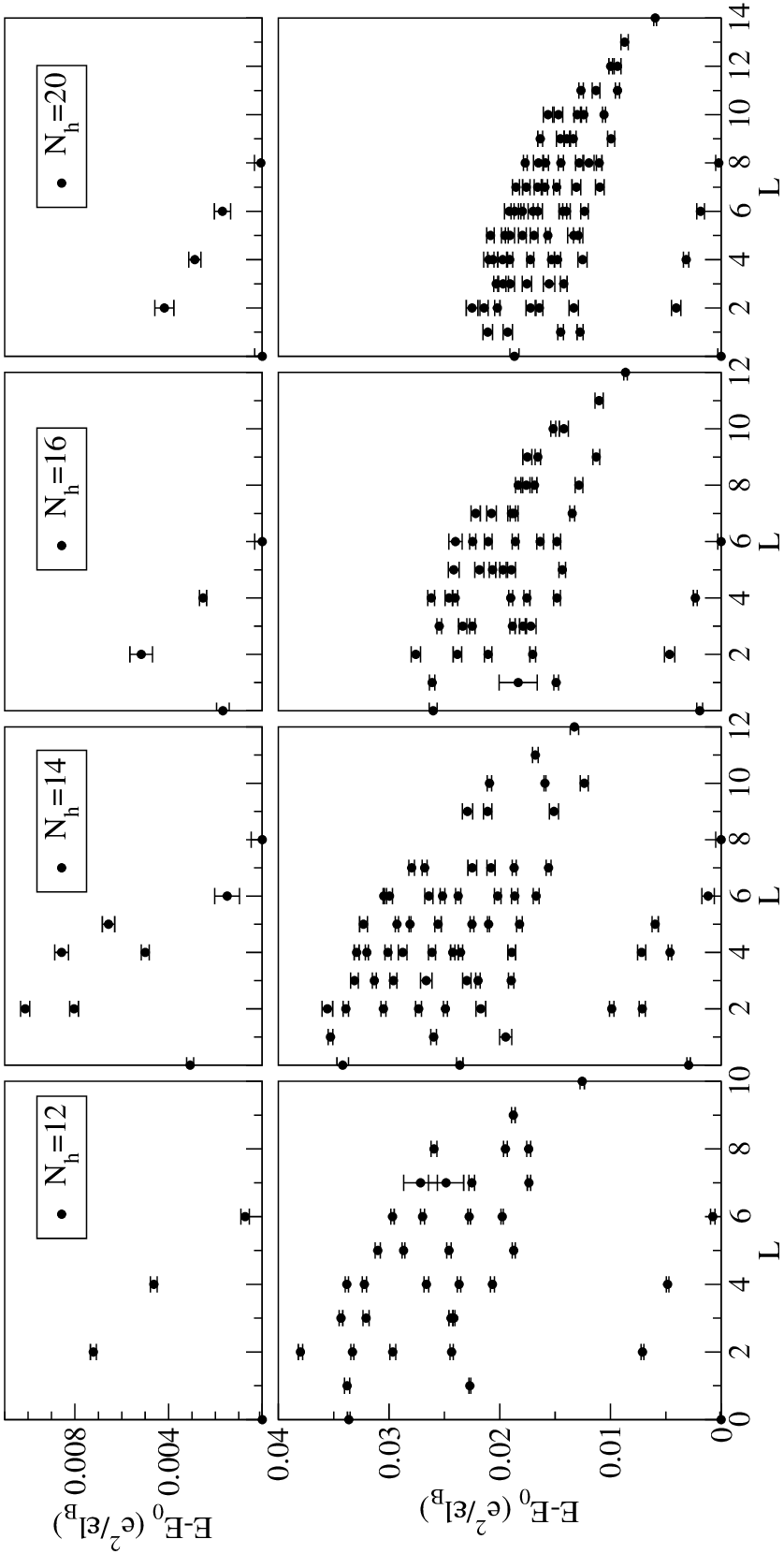}
\end{center}
\caption{\label{speclll}
Zeroth-order (top) and first-order (bottom) CF diagonalization excitation spectra for $N_h=12,14,16,20$ holes
in the lowest Landau level.
}
\end{figure*}

Figure \ref{speclll} shows analogous results for the half filled {\em lowest} LL.   
The zeroth order CF diagonalization generates the lowest band, and the first order 
generates the next band.   The energies of states in the lowest band do not 
change appreciably from zeroth to the first order. 
The energy gap between the two lowest bands can be understood as the energy cost
of exciting one more CF particle-hole pair.  No such bands are seen for half 
filled second LL.

It is not known how this description of the 5/2 FQHE relates to the Pfaffian model.  
In particular, a natural description of the quasiparticles is as excited composite fermions
(which are heavily renormalized by the residual interaction).  From this perspective, 
there is no reason to suspect that they would obey nonabelian statistics, although that 
cannot be ruled out as the residual interaction causes nonperturbative change.

\section{Acknowledgment}

We thank IDRIS-CNRS for a computer time allocation, 
the High Performance Computing (HPC) group at Penn State University ASET 
(Academic Services and Emerging Technologies)
for assistance and computing time on the Lion-XL and Lion-XO clusters. 
Partial  support of this research by the National Science Foundation is gratefully acknowledged.


\newcommand{\PRL}{Phys.\ Rev.\ Lett.}
\newcommand{\PRB}{Phys.\ Rev.\ B}
\newcommand{\PRD}{Phys.\ Rev.\ D}
\newcommand{\NPB}{Nucl.\ Phys.\ B}


\begin{thebibliography}{99}

\bibitem{Willett1} R.~Willett, J.~P.~Eisenstein, H.~L.~Stormer, D.~C.~Tsui, A.~C.~Gossard, and J.~H.~English, \PRL\ \textbf{59}, 1776 (1987).

\bibitem{Pan} W. Pan, J.-S. Xia, V. Shvarts, D. E. Adams, H. L. Stormer, D. C. Tsui, L.
N.  Pfeiffer, K. W. Baldwin, and K. W.  West, Phys. Rev. Lett. {\bf 83}, 3530 (1999).

\bibitem{Eisen02} J.~P.~Eisenstein, K.~B.~Cooper, L.~N.~Pfeiffer, and K.~W.~West, \PRL\ \textbf{88}, 076801 (2002).

\bibitem{FStheory} V.~Kalmeyer and S.~C.~Zhang, \PRB\ \textbf{46}, R9889 (1992);
B.~I.~Halperin, P.~A.~Lee, and N.~Read, \PRB\ \textbf{47}, 7312 (1993).

\bibitem{FSexp} R.~L.~Willett, R.~R.~Ruel, K.~W.~West, and L.~N.~Pfeiffer, \PRL\ \textbf{71}, 3846 (1993); W.~Kang, H.~L.~Stormer, L.~N.~Pfeiffer, K.~W.~Baldwin, and K.~W.~West, \PRL\ \textbf{71}, 3850 (1993);
V.~J.~Goldman, B.~Su, and J.~K.~Jain, \PRL\ \textbf{72}, 2065 (1994);
J.~H.~Smet {\em et al.}, \PRL\ \textbf{77}, 2272 (1996).

\bibitem{Moore1} G.~Moore and N.~Read, \NPB\ \textbf{360}, 362 (1991).

\bibitem{GWW1} M.~Greiter, X.~G.~Wen, and F.~Wilczek, \PRL\ \textbf{66}, 3205 (1991); \NPB\ \textbf{374}, 567 (1992).

\bibitem{Morf1} R.~H.~Morf, \PRL\ \textbf{80}, 1505 (1998).

\bibitem{Park1} K. Park, V. Melik-Alaverdian, N.E. Bonestel, and 
J.K. Jain, Phys. Rev. B {\bf 58}, R10167 (1998).

\bibitem{Scarola1} V.~W.~Scarola, K.~Park, and J.~K.~Jain, Nature \textbf{406}, 863 (2000).

\bibitem{Rezayi00} E.~H.~Rezayi and F.~D.~M.~Haldane, \PRL\ \textbf{84}, 4685 (2000).

\bibitem{Scarola2} V.W. Scarola, J.K. Jain, E.H. Rezayi, Phys. 
Rev. Lett. {\bf 88}, 216804 (2002).

\bibitem{Nayak96} C.~Nayak and F.~Wilczek, \NPB\ \textbf{479}, 529 (1996).

\bibitem{Read96} N.~Read and E.~H.~Rezayi, \PRB\ \textbf{54}, 16864 (1996).

\bibitem{TSimon} Y.~Tserkovnyak and S.~H.~Simon, \PRL\ \textbf{90}, 016802 (2003).

\bibitem{DasSarma05} S.~Das Sarma, M.~Freedman, and C.~Nayak, \PRL\ \textbf{94}, 166802 (2005).

\bibitem{Stern06} A.~Stern and B.~I.~Halperin, \PRL\ \textbf{96}, 016802 (2006).

\bibitem{Bonderson06a} P.~Bonderson, A.~Kitaev, and K.~Shtengel, \PRL\ \textbf{96}, 016803 (2006).

\bibitem{Freedman06} M.~Freedman, C.~Nayak, and K.~Walker, \PRB\ \textbf{73}, 245307 (2006).

\bibitem{Bonderson06b} P.~Bonderson, K.~Sthengel, and J.~K.~Slingerland, \PRL\ \textbf{97}, 016401 (2006).

\bibitem{Bena06} C.~Bena and C.~Nayak, \PRB\ \textbf{73}, 133335 (2006).

\bibitem{Toke07} C.~T\H oke, N.~Regnault, and J.~K.~Jain, \PRL {\bf 98}, 036806 (2006).

\bibitem{Toke06} C.~T\H oke and J.~K.~Jain, \PRL\ \textbf{96}, 246805 (2006).

\bibitem{Haldane83} F.~D.~M.~Haldane, \PRL\ \textbf{51}, 605 (1983); also in \textit{The Quantum Hall Effect}, edited by R.E. Prange and S.M.~Girvin (Springer, New York, 1987).

\bibitem{Fano} G.~Fano, F.~Ortolani, and E.~Colombo, \PRB\ \textbf{34}, 2670 (1986).


\bibitem{Jain1} J.~K.~Jain, \PRL\ \textbf{63}, 199 (1989); Physics Today \textbf{53}(4), 39 (2000).

\bibitem{LM} J.M. Leinaas and J. Myrheim, Nuovo Cimento B {\bf 37}, 1 (1977).

\bibitem{Halperin84} B.I. Halperin, Phys. Rev. Lett. {\bf 52}, 1583
(1984).

\bibitem{Arovas} D.P. Arovas, J.R. Schrieffer, and F. Wilczek,
Phys. Rev. Lett. {\bf 53}, 722 (1984).

\bibitem{Kjonsberg} H.~Kj{\o}nsberg and J.~Myrheim, Int.\ J.\ Mod.\ Phys.\ A \textbf{14}, 537 (1999); H. Kj{\o}nsberg and J.M. Leinaas, Nucl. 
Phys. B {\bf 559}, 705 (1999).

\bibitem{Jeon03} G.S. Jeon, K.L. Graham, J.K. Jain, Phys. Rev. Lett. {\bf 91}, 036801 (2003); Phys. Rev. B {\bf 70},  125316 (2004).

\bibitem{GWW2} M.~Greiter, X.~G.~Wen, and F.~Wilczek, \PRB\ \textbf{46}, 9586 (1992).

\bibitem{JainKamilla} J.~K.~Jain and R.~K.~Kamilla, Int.\ J.\ Mod.\ Phys.\ \textbf{B11}, 2621 (1997); \PRB\ \textbf{55}, R4895 (1997).

\bibitem{Mandal00} S.~S.~Mandal and J.~K.~Jain, \PRB\ \textbf{66}, 155302 (2002).

\bibitem{Morf2} R.~H.~Morf, N.~d'Ambrumenil, and S.~Das Sarma, \PRB\ \textbf{66}, 075408 (2002).

\end{thebibliography}
\end{document}